\documentclass[aps,prd,preprint,superscriptaddress,tightenlines,nofootinbib,showpacs]{revtex4}

\usepackage{graphicx}
\usepackage{dcolumn}
\usepackage{bm}

\begin{document}

\preprint{\vbox{\hbox{\hfil CLNS 06/1984}\hbox{\hfil CLEO 06-24}
}}

\title{Comparison of particle production in quark and gluon fragmentation at $\sqrt{s}\sim$10 GeV}

\author{R.~A.~Briere}
\author{T.~Ferguson}
\author{G.~Tatishvili}
\author{H.~Vogel}
\author{M.~E.~Watkins}
\affiliation{Carnegie Mellon University, Pittsburgh, Pennsylvania 15213}
\author{J.~L.~Rosner}
\affiliation{Enrico Fermi Institute, University of
Chicago, Chicago, Illinois 60637}
\author{N.~E.~Adam}
\author{J.~P.~Alexander}
\author{D.~G.~Cassel}
\author{J.~E.~Duboscq}
\author{R.~Ehrlich}
\author{L.~Fields}
\author{R.~S.~Galik}
\author{L.~Gibbons}
\author{R.~Gray}
\author{S.~W.~Gray}
\author{D.~L.~Hartill}
\author{B.~K.~Heltsley}
\author{D.~Hertz}
\author{C.~D.~Jones}
\author{J.~Kandaswamy}
\author{D.~L.~Kreinick}
\author{V.~E.~Kuznetsov}
\author{H.~Mahlke-Kr\"uger}
\author{P.~U.~E.~Onyisi}
\author{J.~R.~Patterson}
\author{D.~Peterson}
\author{J.~Pivarski}
\author{D.~Riley}
\author{A.~Ryd}
\author{A.~J.~Sadoff}
\author{H.~Schwarthoff}
\author{X.~Shi}
\author{S.~Stroiney}
\author{W.~M.~Sun}
\author{T.~Wilksen}
\author{M.~Weinberger}
\author{}
\affiliation{Cornell University, Ithaca, New York 14853}
% ...
\collaboration{CLEO Collaboration} 
\noaffiliation

\date{April 20, 2007}

\begin{abstract}
Using data collected with the CLEO~III detector at the Cornell Electron Storage
Ring, 
we study the inclusive production of baryons/antibaryons (p, $\Lambda$) and
mesons ($\phi$ and $f_2(1270)$) in gluon-fragmentation and quark-fragmentation
processes.  We first corroborate
previous per-event total particle
yields in $\Upsilon$(1S)$\to ggg$
compared with nearby continuum ($e^+e^-\to q{\overline q}$)
indicating
greater ($\sim\times$2) per-event yields of baryons in 3-gluon
fragmentation. We find similar
results when we extend that comparison to include the
$\Upsilon$(2S) and $\Upsilon$(3S) resonances. With higher
statistics, we now also probe the momentum dependence
of these per-event particle yields.
Next, we compare
particle production in the photon-tagged process
$\Upsilon({\rm 1S}) \rightarrow gg\gamma$ with that
in $e^+e^- \rightarrow q{\bar q}\gamma$ events, to allow
comparison of two-parton with three-parton
particle-specific fragmentation. 
For each particle,
we determine the `enhancement' ratio, defined
as the ratio of
particle yields per gluon
fragmentation event compared
to quark fragmentation event. Thus
defined, an enhancement of 1.0 implies
equal per-event production in gluon and quark 
fragmentation.
In the photon-tagged analysis ($\Upsilon({\rm 1S}) \rightarrow gg\gamma$ compared to
 $e^+e^- \rightarrow q{\bar q}\gamma$),
we find almost no enhancement for protons ($\sim$ 1.2 $\pm$ 0.1), 
but a significant enhancement ($\sim$ 1.9 $\pm$ 0.3) for
$\Lambda$'s.
This small measured proton enhancement rate is supported by a study
of baryon production in $\chi_{b2}\to gg\to p+X$ relative
to $\chi_{b1}\to q{\overline q}g\to p+X$. 
Overall, per-event
baryon production in radiative two-gluon fragmentation is somewhat
smaller than that observed in
three-gluon decays of the $\Upsilon$(1S). 
Our results for baryon
production are inconsistent with the predictions of the JETSET (7.3)
fragmentation model.
\end{abstract}

\newpage

\maketitle

\tighten

{
 \renewcommand{\thefootnote}
 {\fnsymbol{footnote}}
 \setcounter{footnote}{0}
}

\newpage

\setcounter{footnote}{0}

\section{Introduction}
Understanding
hadronization, the process by which elementary partons (gluons and quarks)
evolve into mesons and baryons, 
is complicated by its intrinsically
non-perturbative nature. 
Due to the fact that gluons carry two color indices whereas
quarks carry only one, the intrinsic
gluon-gluon coupling strength ($C_A$=3) is larger than the
intrinsic quark-gluon coupling strength ($C_F$=4/3).
Radiation of secondary
and tertiary gluons is therefore expected to be
more likely when hadronization is initiated by
a gluon rather than by a quark. This results in a greater number of 
final state hadrons as well as a larger average transverse momentum
in the former case compared to the latter case. In the limit
$Q^2\to\infty$, the ratio of the number of hadrons produced in
gluon-initiated jets to the number of
hadrons produced in quark-initiated jets is expected, in
lowest order, to approach a simple
color-counting ratio 
9/4\cite{r:theory76}.

Many experiments have searched for, and found,
multiplicity and jet shape differences between quark and
gluon fragmentation.
At $Z^0$ energies, {\it $q{\overline q}$g} events
are distinguished by their three-jet topology. Within such events,
quark and
gluon jets can be separated by a variety of techniques including vertex 
tagging.  Because 
gluons rarely fragment into heavy quarks, they will produce jets that 
form a vertex
at the $e^+e^-$ interaction
point. Quark jets, to the contrary, tend to form a detached
vertex when the jet contains
a long-lived bottom or charm quark. For light-quark events with
gluon radiation, however,
the assignment of final state hadrons to the initial
state partons is generally more ambiguous and often relies on Monte Carlo simulations
to determine the fraction of times that an observed hadron is correctly
traced to a primary parton.
At lower energies, one can exploit the decay characteristics of
quarkonium states to directly compare gluon and quark fragmentation
using data taken both on-resonance and off-resonance (on the continuum),
respectively.
The 10 GeV center of mass energy range offers a unique 
opportunity to
probe quark and gluon fragmentation effects, without relying on Monte
Carlo simulation to associate the final state hadrons with an initial state 
parton. 
CLEO\cite{r:CLEO91} found that
the thrust and charged multiplicity distributions of
$\chi_{b0}$ and $\chi_{b2}$ two-gluon decays are more similar
to $\Upsilon$(1S)$\to ggg$ than to
continuum $e^+e^-\to q{\overline q}$ events; the
reverse was found to be true 
for $\chi_{b1}\to q{\overline q}g$.

Specific particle production in gluon- and quark-fragmentation
has also been studied.
Within the limits of their precision, previous studies at
SLD found inclusive production of pions, kaons and protons to
be equivalent for gluon-tagged and quark-tagged jets\cite{r:SLD01}.
OPAL has measured inclusive charm production to be
$(3.20\pm0.21\pm0.38)$\% in gluon jets\cite{r:OPALincl04,r:OPALcharm99},
more than an order of magnitude smaller than the rate observed
in quark jets at the $Z^0$. ALEPH\cite{r:ALEPHb99} and DELPHI\cite{r:DELPHIb99}
both measured inclusive bottom production in gluon-tagged jets
to be $2-3\times 10^{-3}$, again considerably smaller than
that expected from charge counting in quark fragmentation.
Most directly comparable to our current work, OPAL has
also compared inclusive $K^0_{\rm s}$ and $\Lambda$ production
in gluon- vs. quark-tagged jets in 
$e^+e^-\to q{\overline q}g$ events, finding inclusive production
ratios ($g/q$) consistent with unity ($0.94\pm0.07\pm0.07$ and
$1.18\pm0.01\pm0.17$, respectively)\cite{r:OPAL98K0slam}.

The decay 
$\Upsilon({\rm 1S}) \rightarrow gg\gamma$ allows one to 
directly compare the $gg$ system 
in a $gg\gamma$ event with the $q{\bar q}$ system in  
$e^+e^- \rightarrow q{\bar q}\gamma$ events. In these cases,
the system recoiling against the photon consists (to lowest order) of hadrons
that have evolved from either a two-gluon or a quark-antiquark system. 
The properties of the recoil systems can then be compared.\footnote{Although
there may be gluon radiation from the initial partons, we 
do not distinguish such radiation explicitly in this analysis.
Thus, the states that we are comparing are, strictly speaking,
$gg\gamma$ and $q{\bar q}\gamma$ to lowest-order only; 
additional gluon radiation, to which we are not experimentally sensitive,
may be present in many of the events in our sample.
Without the
ability to adequately identify additional gluons, such
higher-order radiative
effects are therefore implicitly absorbed into the experimental 
measurement.}
Additionally, the radiative transitions from the radially excited
$\Upsilon$ states to the orbitally excited $\chi_b$ triplet offer
an opportunity to further probe fragmentation differences between
decays of the J=0 and J=2 $\chi_b$ states, which decay predominantly
to two gluons, vs. decays of the J=1 state. Since the J=1 state
is prohibited from decaying into two on-shell gluons, the decay into
one hard and one soft, nearly on-shell
virtual gluon ($gg^*$, followed by $g^*\to q{\overline q}$) is kinematically
most favored. 
Statistical
correlations
between transition photons with
inclusive production of particular final-state
particles ($X$) allows a measurement of the relative yields of
$gg\to X:q{\overline q}(g)\to X$ to these species.

In an over-simplified `independent fragmentation' model,
hadronization occurs independently for each parton. 
In such a picture, if fragmentation of each
parton (gluon or quark) of a given energy
is identical, then the ratio
of particle production for $gg\gamma:q{\overline q}\gamma:(\chi_b\to gg):ggg$
hadronization
should vary as: 2:2:2:3.
In the opposite extreme, fragmentation occurs in the
stretching `strings' between the two partons, in
which case the above ratio should be 1:1:1:3. 

In this analysis, we focus on the relative production rates of
baryons ($p$ and $\Lambda\to p\pi$) 
and heavy mesons ($\phi\to K^+K^-$ and $f_2(1270)\to\pi^+\pi^-)$ in gluon vs.
quark fragmentation (charge conjugation is implied). 
A previous study noted enhancements in
the production of $\phi$, $\Lambda$ and $p$ in three-gluon 
decays of the $\Upsilon$(1S)\cite{r:cleo84}, at a statistical significance 
of no more than 
than 2-3~$\sigma$. That initial study also
found approximately one unit larger 
charged multiplicity for three-gluon
fragmentation of the $\Upsilon$(1S) 
compared to $q{\overline q}$ fragmentation at
a comparable center-of-mass energy. With the limited statistics 
available at that
time, the additional unit of multiplicity could entirely be accounted for
by enhanced three-gluonic production of baryons.
We now have sufficient
statistics to re-measure the three-gluon particle production rates,
and also to compare, for the first time, 
inclusive production in two-gluon fragmentation vs.
inclusive production in three-gluon fragmentation. 

Since then, other experimental data on quark/gluon fragmentation differences
in the $\sqrt{s}\sim$10 GeV energy regime
have become available, including:
\begin{enumerate}
\item The observation that fragmentation of the J=1 state of the $\chi_b$ triplet 
($\chi_{b1}\to q{\overline q}$g$\to charm$) 
results in charm production comparable to the 
underlying continuum; no such charm production 
is observed in the two-gluon decays of the J=0 or J=2 states\cite{r:ichep06}.
\item An enhancement in production of hidden charm in gluonic decays of
the $\Upsilon$ resonances:
$(\Upsilon$(1S)$\to ggg\to J/\psi+X)/(e^+e^-\to J/\psi+X)\gtrsim$5\cite{CLEOpsi04} at 90\% c.l.
\item Production of deuterons from resonant 3-gluon 
decays of both the 
$\Upsilon$(1S) and $\Upsilon$(2S) at the level of $10^{-3}$; 
no significant
production of deuterons is observed from the continuum\cite{CLEOdeuteron06}.
Enhancements per event are $\ge$10.
\item Production of $\eta'$ in gluonic decays of
the $\Upsilon$ resonance of similar magnitude to that observed in
$\Upsilon$ decays via $q{\overline q}$:
$(\Upsilon$(1S)$\to ggg\to \eta'+X)/(\Upsilon\to {q\overline q}\to \eta'+X)\sim$2/3),
integrated over momentum\cite{CLEOetaprime02}.
\end{enumerate}

\section{Detector and Data Sample}
The CLEO~III detector\cite{r:CLEOIIIa,r:CLEOIIIb,r:CLEOIIIc} 
is a general purpose solenoidal magnet spectrometer and
calorimeter.  The main components of the detector used in this analysis 
are the drift chamber and the silicon detector used for track finding, the
crystal calorimeter for energy measurements, and the Ring Imaging Cherenkov 
detector (RICH) and specific ionization loss in the drift chamber for 
particle identification.
This system is very efficient ($\epsilon\ge$98\%) 
for detecting tracks that have transverse momenta ($p_T$)
relative to the
beam axis greater than 200 MeV/c, and that are contained within the good
fiducial volume of the drift chamber ($|\cos\theta|<$0.93, with $\theta$
defined as the polar angle relative to the beam axis). Below this 
threshold, the charged particle detection efficiency in the fiducial
volume decreases to 
approximately 90\% at $p_T\sim$100 MeV/c. For $p_T<$100 MeV/c, the efficiency
decreases roughly linearly to zero at a threshold of $p_T\approx$30 MeV/c.
Just within the solenoidal magnet coil is the electromagnetic calorimeter,
consisting of 7800 thallium doped CsI crystals.  The central region
of the calorimeter covers about three-quarters of the solid angle and has an energy
resolution of
\begin{equation}
\frac{ \sigma_{\rm E}}{E}(\%) = \frac{0.6}{E^{0.73}} + 1.14 - 0.01E,
                                \label{eq:resolution1}
\end{equation}
with $E$ the shower energy in GeV. This parameterization translates to an
energy resolution of about 2\% at 2 GeV and 1.2\% at 5 GeV. Two end-cap
regions of the crystal calorimeter extend solid angle coverage to about 95\%
of $4\pi$, although energy resolution is not as good as that of the
central region. 
The tracking system, RICH particle identification system and calorimeter
are all contained 
within the 1.5 Tesla superconducting coil. 
Flux return and tracking
chambers used for muon detection are located immediately outside the coil and 
in the two end-cap regions.

We use the CLEO-III
data collected at the narrow $\Upsilon$
resonances as a source of $ggg$ and 
$gg\gamma$ events, and data taken just
below the narrow resonances, as well as the below-4S continuum
($\sqrt{s}$=10.55 GeV)
as a source of $q{\bar q}$ and
$q{\bar q\gamma}$ events. 
Since $\Upsilon$(4S)$\to B{\overline B}\sim$100\%,
data collected on the broad $\Upsilon$(4S) resonance is analyzed as
a `control' sample, for which we expect no deviation from 
the below-4S continuum when we require a photon having
$z_\gamma=E_\gamma/E_{\rm beam}>$0.5. 

The $\gamma$ in our continuum
$q{\bar q\gamma}$ sample results primarily from
initial state radiation (ISR)\cite{r:bkqed}.
We compare events for which the fractional photon
energies
are the same, which
ensures that the recoil systems (either two-gluon or
$q{\overline q}$) have comparable energies. This convention
deviates slightly
from that of our previous publication\cite{r:lauren-paper} 
for which the scaling variable
was the
recoil mass of the $gg$ and $q{\overline q}$ 
systems opposite the hard photon 
($M_{recoil}$, defined by $M_{recoil}$ =
$\sqrt{4E_{\rm beam}^{2}(1-E_{\gamma}/E_{\rm beam})}$).
Comparison with
continuum data taken $\sim$20 MeV below each of the $\Upsilon$
resonances mitigates the effect of the $\sim$1 GeV continuum
center-of-mass
energy extrapolation between the $\Upsilon$(1S) and
below-4S data samples required in the previous analysis\cite{r:cleo84}, for
which continuum data were only taken in the 10.55 GeV center-of-mass region.
To compare $ggg$ with $q{\overline q}$ hadronization, we simply
bin by scaled momentum of the particle in question.

\subsection{Event Selection}
We impose event-selection requirements identical to those used
in our previous study of inclusive direct photon production in 
$\Upsilon$ decays\cite{r:shawn}.
Those cuts are designed primarily to suppress backgrounds such as two-photon collisions,
QED events (including tau pair production), and beam-gas and beam-wall collisions.  
Luminosity,
event count, and photon yields ($z_\gamma>$0.5) are
given in Table \ref{tab:datasets}.
\begin{table}[htpb]
\caption{\label{tab:datasets}
Summary of data and JETSET Monte Carlo used in analysis. 
For each data set, we track the number of photons per unit
luminosity, as well as the total number of observed 
hadronic events per unit luminosity ${\cal L}$.  
HadEvts denotes the total 
number of events in each sample identified as hadronic by our 
event selection requirements. The number of photons
having scaled momentum $z_\gamma$ 
greater than 0.5
is presented in the last column. For $B{\overline B}$ Monte Carlo
simulations, the small number of observed high-energy photons is
a result of detector resolution and mis-reconstruction.}
\begin{tabular}{c|c|c|c|c|c|c} 
Data Type & Type & Resonance & E$_{\rm cm}$ (GeV) & ${\cal L}$ (${\rm pb}^{-1}$) & HadEvts 
($\times 10^3$)&
$N_\gamma(z>0.5)$ ($\times 10^2$)\\ 
\hline 
1S & Data & $\Upsilon$(1S) & 9.455-9.465 & 1220  & 22780 & 2190 \\
2S & Data & $\Upsilon$(2S) & 10.018-10.028 & 1070  & 9450 & 888 \\
3S & Data & $\Upsilon$(3S) & 10.350-10.360 & 1420  & 8890 & 795 \\
4S & Data & $\Upsilon$(4S) & 10.575-10.585 & 5520  & 18970 & 1650 \\
1S-CO & Data & $<\Upsilon$(1S) & 9.400-9.454 & 144  & 515 & 57 \\
2S-CO & Data & $<\Upsilon$(2S) & 9.523-10.017 & 312  & 932 & 103 \\
3S-CO & Data & $<\Upsilon$(3S) & 10.083-10.349 & 185  & 532 & 59 \\
4S-CO & Data & $<\Upsilon$(4S) & 10.410-10.574 & 2100  & 5680 & 647 \\ 
1S & JETSET MC & $\Upsilon$(1S) & 9.455-9.465 &  & 1160 & 99 \\
2S & JETSET MC & $\Upsilon$(2S) & 10.018-10.028 &  & 9190 & 700 \\
3S & JETSET MC & $\Upsilon$(3S) & 10.350-10.360 &  & 3890 & 270 \\
4S & B\=B MC & $\Upsilon$(4S) & 10.575-10.585 &  & 8350 & 3 \\
1S-CO & JETSET MC & $<\Upsilon$(1S) & 9.400-9.454 &  & 8170 & 681 \\
2S-CO & JETSET MC & $<\Upsilon$(2S) & 9.523-10.017 &  & 7610 & 666 \\
3S-CO & JETSET MC & $<\Upsilon$(3S) & 10.083-10.349 &  & 12850 & 1150 \\
4S-CO & JETSET MC & $<\Upsilon$(4S) & 10.410-10.574 &  & 63630 & 5680 \\ 
\hline
\end{tabular} 
\end{table}

\subsection{Background Suppression}
To determine the characteristics of resonant
$\Upsilon\to gg\gamma$ events, 
we must subtract the background arising from non-resonant
$q{\bar q\gamma}$ and 
$e^+e^-\to\tau\tau\gamma$ events produced in continuum
$e^+e^-$ annihilations at
$\sqrt{s}=M_{\Upsilon({\rm nS})}$, with n=1, 2, or 3. 
This is done by direct scaling of the event samples collected 
off-resonance on the nearby continuum.

\begin{figure*}
\includegraphics[width=6.4in,height=3.2in]{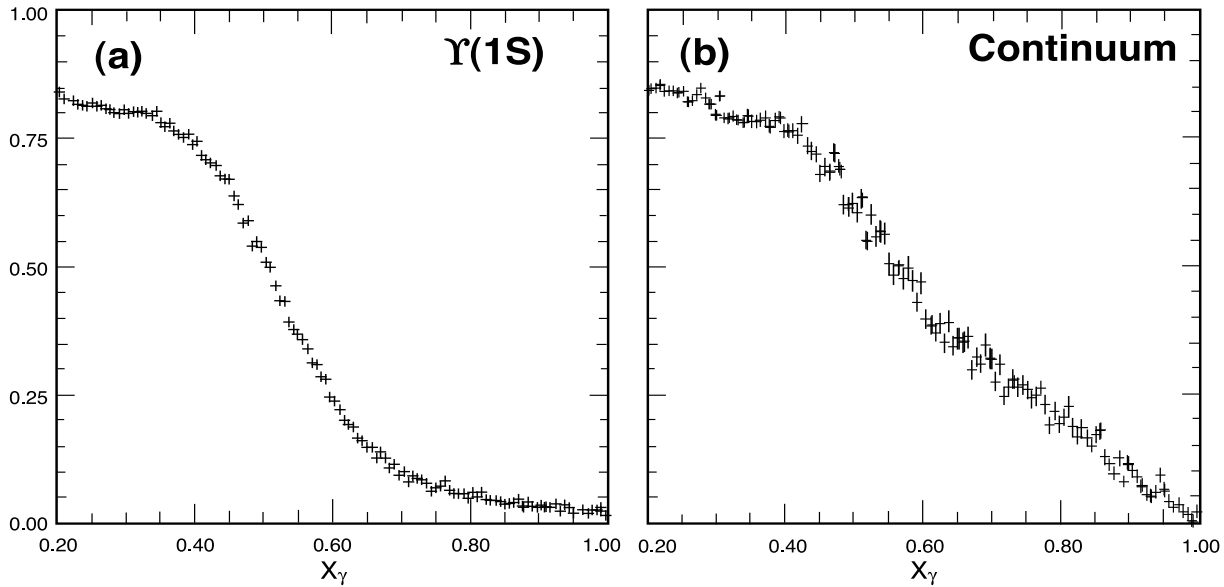}
     \caption{\label{fig:pi01s} JETSET Monte
Carlo prediction for fraction of photons not produced directly,
but through the decay of neutral particles 
(such as $\pi^0$, $\eta$, $\eta$', and $\omega$)
on the 1S resonance (left) and on the continuum below
the 4S resonance (right). }
\end{figure*}

In order to isolate continuum
$q{\bar q\gamma}$ events, 
$\tau\tau\gamma$ contamination must be explicitly subtracted,
using a Monte Carlo simulation of tau pair events. 
We find that  
$\tau\tau\gamma$ events comprise about 5\% of the $q{\bar q\gamma}$
data sample passing the event selection cuts\cite{r:shawn}.
Beam-gas and two-photon backgrounds were investigated and found 
to be negligibly small.
The photon-tagged sample can also be contaminated by cases where
the high-energy photon candidate
is not produced directly, but is
actually either a secondary daughter
(mostly from $\pi^0$ decay) or a mis-identified hadronic
shower.
Figure~\ref{fig:pi01s} illustrates the fraction of photons in
Monte Carlo simulations of 
on-1S resonance and below-4S continuum, respectively, that are
not produced in a direct decay.

Integrated over all tag photon momenta considered
in this analysis, $\pi^0$ contamination comprises a
$\sim$15\% background to the direct-photon sample.
Monte Carlo simulations also indicate that the $\pi^0$ 
contamination tends to cancel when we take ratios of resonant photon
production to continuum photon production.

\subsubsection{Particle Identification}
Our photon and particle identification procedures are 
identical to those developed in \cite{r:shawn}.
Photon candidates are selected from showers 
with widths and patterns of energy deposition consistent with those of a 
photon, as opposed 
to a neutral hadron
(e.g., $\pi^0$ with overlapping photon showers, $K^0_{\rm L}$, neutrons, etc.). 
To ensure that the 
events are well-contained within the CLEO detector,
we require 
$|\cos\theta_\gamma|<0.707$ ($\theta_\gamma$ defined as before
as the polar angle between the beam axis and the direct photon).
For $p$ (and $\overline{p}$), we require that charged tracks have specific
ionizatation (dE/dx) and also RICH information
consistent with those expected for protons. For momenta less
than 1 GeV/c, we also require that the
associated charged track dE/dx 
information be
inconsistent (at the two standard-deviation level, with 
$\sigma$ the momentum-dependent specific ionization
resolution) with that expected for true
pions. Although this results in a discontinuity in particle
identification efficiency at 1 GeV/c, this requirement is
necessary to ensure a high-purity sample.
For all $p$ and $\overline{p}$ candidates, 
we require that $p$ ($\overline{p}$)
momenta exceed 400 MeV/c to 
suppress beam-wall and fake backgrounds 
(i.e. K$^+$ and $\pi^+$ that pass $p$ identification cuts) 
and also to eliminate concerns
regarding protons ranging out in the beampipe.
For reconstruction of 
$\phi$ ($f_2$(1270)) from kaons (pions) we require that 
pairs of opposite charged tracks with
momenta greater than 200 MeV/c (500 MeV/c) have particle identification
information consistent with their assumed identities.
$\Lambda$'s are identified using the standard
CLEO algorithms for reconstruction of detached vertices.
Tables \ref{a} and \ref{b} summarize the raw, observed particle yields for our 
measurements, for data and Monte Carlo simulations, respectively.
\begin{table}[htpb]
\begin{tabular}{c|c|c} \hline
Particle Type & (ggg)/($q\overline{q}$) [Data] & (gg$\gamma$)/($q\overline{q} \gamma$)[Data]  \\
\hline
$\Lambda$ & ($873600\pm1400$)/($107300\pm600$) & ($3480\pm90$)/($570\pm60$) \\
$p$ & ($1399800\pm1200$)/($295900\pm500$) & ($7970\pm90$)/($2190\pm50$)  \\
$\overline{p}$ & ($1359500\pm1200$)/($285400\pm500$) & ($7830\pm90$)/($2090\pm50$)  \\
$\phi$ & ($227900\pm1600$)/($48300\pm800$) & ($1950\pm150$)/($380\pm70$) \\ 
$f_2(1270)$ & ($193000\pm4000$)/($66500\pm1800$) & ($1600\pm400$)/($400\pm200$) \\
\hline
\end{tabular} 
\caption{\label{a}
Data particle yields for the on-1S resonance compared to continuum events.  
First column is particle type.  Second and third colunmns show particle 
counts for the data with in the format of (resonance yield) / (continuum yield)
for the three gluon (2nd column) and two gluon one photon (3rd column) analyses.}
\end{table}

\begin{table}[htpb]
\begin{tabular}{c|c|c} \hline
Particle Type & 
(ggg)/($q\overline{q}$) [JETSET MC] & (gg$\gamma$)/($q\overline{q} \gamma$) [JETSET MC] \\
\hline
$\Lambda$ & ($136700\pm500$)/($1333200\pm2000$) & ($690\pm30$)/($6410\pm150$) \\
$p$ & ($266600\pm500$)/($3334200\pm1800$) & ($1650\pm40$)/($20660\pm140$) \\
$\overline{p}$ & ($257300\pm500$)/($3198300\pm1800$) & ($1590\pm40$)/($19880\pm140$) \\
$\phi$ & ($48100\pm900$)/($837000\pm4000$) & ($380\pm80$)/($6000\pm800$) \\
\hline
\end{tabular} 
\caption{\label{b}
Monte Carlo particle yields for the on-1S resonance compared to continuum events.  
First column is particle type.  Second and third colunmns show particle 
counts for the data with in the format of (resonance yield) / (continuum yield)
for the three gluon (2nd column) and two gluon one photon (3rd column) analyses.}
\end{table}

\subsubsection{Backgrounds to the Proton Sample}
\begin{figure*}
\includegraphics[width=6.4in,height=3.2in]{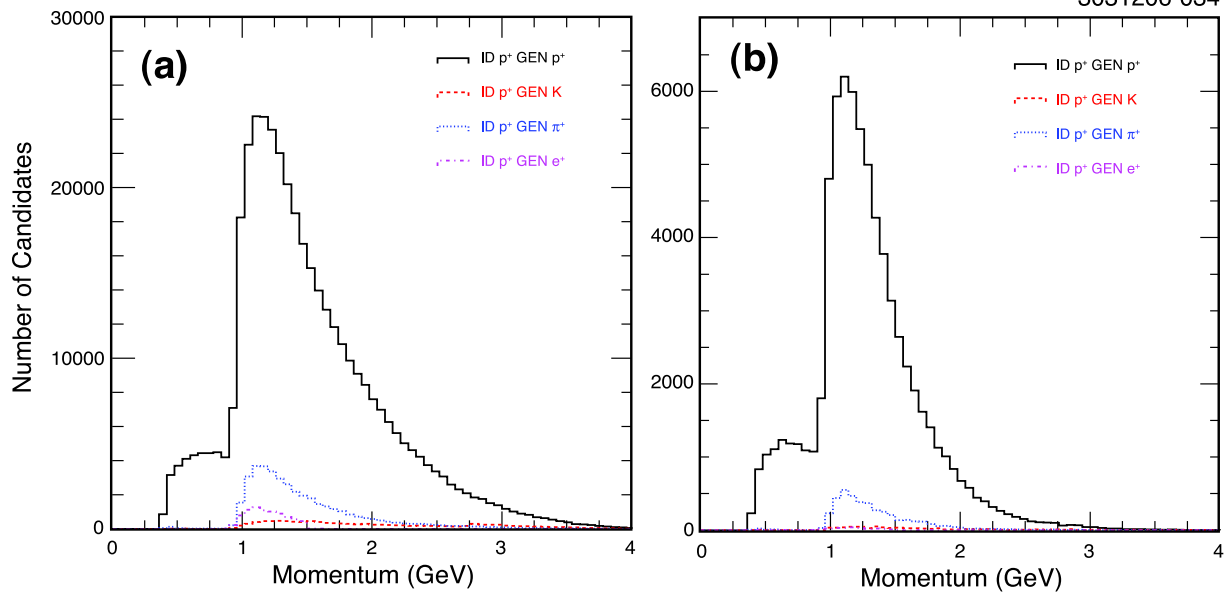}
     \caption{\label{fig:b1sfakes} (Left) Proton fakes for a sample of below-1S
Monte Carlo simulations.  The solid black curve shows the number of all particles
identified as protons that were also tagged as true protons.  The red dashed 
(blue dotted, magenta dash-dot)
curve corresponds to those particles that were identified as protons, but that were
actually kaons (pions, positrons). (Right) Same for on-1S event simulations.
Note the discontinuity at 1 GeV/c, resulting from our momentum-dependent
particle identification requirements below and above that momentum (see text).}
\end{figure*}

We use Monte Carlo simulations to
assess fake proton backgrounds.
Figure~\ref{fig:b1sfakes} illustrates proton fakes for a sample of below-1S
Monte Carlo continuum simulations.  The solid black curve shows the number of all particles
identified as protons that were also tagged as true protons.  The 
red dashed 
(blue dotted, magenta dash-dot)
curve corresponds to those particles that were identified as protons, but that were generated as true kaons (pions, positrons) in the Monte Carlo simulated
event sample.  
Proton backgrounds are 
observed to be present at the $\sim$10\% level and are expected to largely
cancel in the enhancement ratio.

\subsection{Signal Definition}
In this analysis we measure particle enhancements in 
both the $ggg$ and $gg\gamma$
decays of the Upsilon system, relative to $q{\overline q}$($\gamma$) production on the
underlying continuum.  
Our definition of enhancement is given quantitatively
as the continuum-subtracted
resonance yield 
relative to the continuum yield.  
Thus defined, an enhancement of 1 indicates that
a given particle is produced as often (per event) 
on the continuum as on the resonance.
Note that our definition of `continuum' here includes both
continuum below the resonance peak, as well
as resonance$\to q{\overline q}$ through vacuum-polarization;
i.e., all $e^+e^-\to q{\overline q}$-like processes which must be
explicitly subtracted in determining the
characteristics of 3-gluon resonant decays.\footnote{Vacuum polarization
processes are subtracted by direct scaling of the continuum using
the $\Upsilon\to\gamma^\star\to q{\overline q}$ values tabulated
previously\cite{r:shawn}.}
Furthermore, note that
for the $\Upsilon$(2S) and $\Upsilon$(3S) data, there is no subtraction
of cascades to lower $\Upsilon$ states or $\chi_b$ decays. In what follows, 
``$\Upsilon$(2S)'' denotes a sum over $\Upsilon$(2S) direct, 
$\Upsilon({\rm 2S})\to\Upsilon$(1S)+X and $\Upsilon$(2S)$\to\gamma\chi_b$.
Assuming the direct decays of the $\Upsilon$ resonances are identical,
an $\Upsilon$(2S) enhancement smaller than that of the 
$\Upsilon$(1S) implies that the enhancements 
from the first and third processes
enumerated above are therefore smaller than for the $\Upsilon$(1S).

In general we have two
continuum-subtraction options:  we may determine enhancements for all resonances 
relative to the below-4S continuum (for which the statistics are largest, but the 
difference in $e^+e^-$ collision energies is also largest)
or we may find enhancements relative
to their individual 
below-resonant continuua.  For mass-fitted particles
we normalize exclusively to the below-4S contiuum, as the individual continuua
(below-1S, -2S, and -3S) have insufficient
statistics to yield well-fitted
mass peaks.  For particle counts determined by the momentum spectra (protons
and antiprotons), 
we normalize to both the below-4S continuum as well as the
resonance-specific continuua 
and incorporate the differences in the enhancements 
calculated in the two cases into the overall systematic
error.

\subsection{Particle production in three-gluon vs.
$q{\overline q}$ events}
The previous CLEO-I\cite{r:cleo84} 
analysis already observed significant enhancements of
$p$ and $\Lambda$ produced in 3-gluon decays
of the 1S relative to the below-4S continuum.
We repeat that analysis with our larger, current data set, as detailed
below.
Errors on particle yields
are obtained from the error returned
from the fit if the particle count is obtained by 
fitting a mass peak ($\Lambda, \phi, f_2$), or 
by the square root of the total count
if the particle count is obtained from 
a simple integration over the momentum spectrum ($p$, ${\overline p}$).
For the $ggg$ analysis described below, 
we determine enhancements as a function of
scaled momentum and also calculate
momentum-integrated enhancements for each particle, to allow comparison
with previous results.

\subsection{$gg\gamma$ Analysis}

For the $gg\gamma$ analysis we normalize 
the total particle yield to the
photon count in a given photon momentum bin.  
For each bin,
we then find the fractional contamination ${\cal F}$ of resonance photons
``$R$'' 
due to the underlying
continuum ``$C$''
(Eqn. \ref{eq:cntfrac}) in terms of the visible cross-section
$\sigma$
for high-energy photons and the known beam energies ``$E$''.  
\begin{equation}
{\cal F}=\frac{\sigma_{z_\gamma>0.5}^C}{\sigma_{z_\gamma>0.5}^R}\left(\frac{E^C}{E^R}\right)^2
\label{eq:cntfrac}
\end{equation}
Once $\cal F$ is known, the resonance
yield can be extracted by straight-forward algebra.

\section{Results from Upsilon Decays}

\subsection{ggg Enhancements with respect to $q{\overline q}$}

\subsubsection{Baryon Enhancements}

\begin{figure*}
     \includegraphics[width=3.5in,height=3.5in]{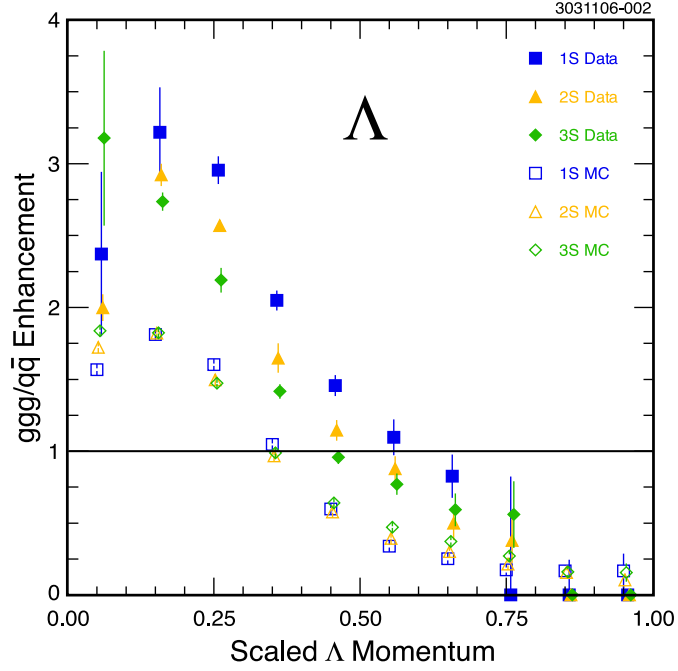}
     \caption{\label{fig:ggglam} Raw
(i.e., observed, and with
no relative efficiency corrections applied)
enhancements for $ggg\to\Lambda+X$ binned
according to scaled momentum ($p_\Lambda/E_{\rm beam}$). 
Blue square (gold triangle, green diamond) symbols correspond to 
enhancements on the 1S (2S, 3S) resonance.  Closed symbols are data, 
open symbols are derived from JETSET 7.3 Monte Carlo simulations. 
No relative efficiency corrections have been applied to these 'raw'
data.}
\end{figure*}

Figure~\ref{fig:ggglam} presents our $\Lambda$ enhancements binned according to scaled
momentum, defined
as before as the momentum of the particle divided by the beam energy.
In the figure, blue square (gold triangle, green diamond) symbols correspond to enhancements on the 1S
(2S, 3S) resonance.  Closed symbols 
are data and open symbols are JETSET 7.3\cite{JETSET7.4}
event generator simulations followed by the full CLEOIII
GEANT-based\cite{r:GEANT} Monte Carlo detector simulation.
From the figure we see that the $\Upsilon$'s show 
qualitatively the same behavior for all resonances (1S, 2S, 3S)
in both data and Monte Carlo, namely a smooth decrease 
in enhancement with increasing scaled momentum.
We note that the enhancements decrease steadily as one goes from 
$\Upsilon$(1S) to 
$\Upsilon$(2S) to 
$\Upsilon$(3S) and that the data, at all scaled momenta, 
show significantly greater enhancements than do the Monte Carlo simulations.

\begin{figure*}
\includegraphics[width=6.4in,height=3.2in]{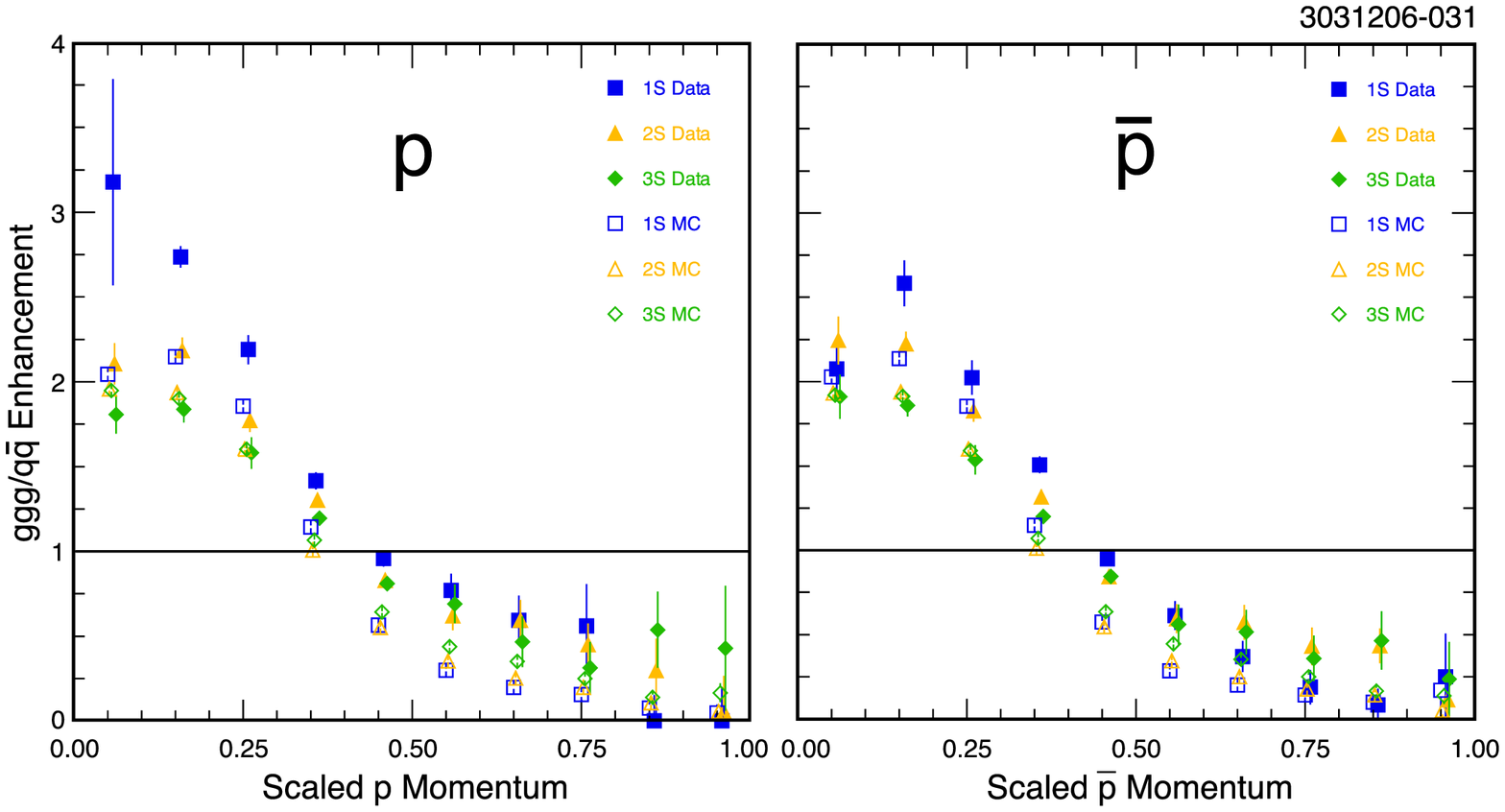}
     \caption{\label{fig:gggpro} (Left) 
Raw enhancements for $ggg\to p+X$ binned
according to scaled momentum.  
Blue square (gold triangle, green diamond) symbols correspond to 
enhancements on the 1S (2S, 3S) resonance.  Closed symbols are data, 
open symbols are JETSET Monte Carlo. (Right) Same for $\overline{p}$.}
\end{figure*}

Figure~\ref{fig:gggpro} 
shows the $p$ and $\overline{p}$ enhancements.
With the exception of the very lowest momentum
bin, which is most
subject to range-out effects,
the consistency between the two indicates 
that beam-wall and beam-gas backgrounds (which produce 
an excess of $p$ in the beam) are not substantial.  As compared
to $\Lambda$ enhancements, $p$ and ${\overline p}$ enhancements are
lower and the differences between 1S, 2S, and 3S enhancements 
(as well as the differences between data and Monte Carlo) are smaller.

\subsubsection{$\phi$ and $f_2$(1270) Enhancements}

\begin{figure*}
\includegraphics[width=6.4in,height=3.2in]{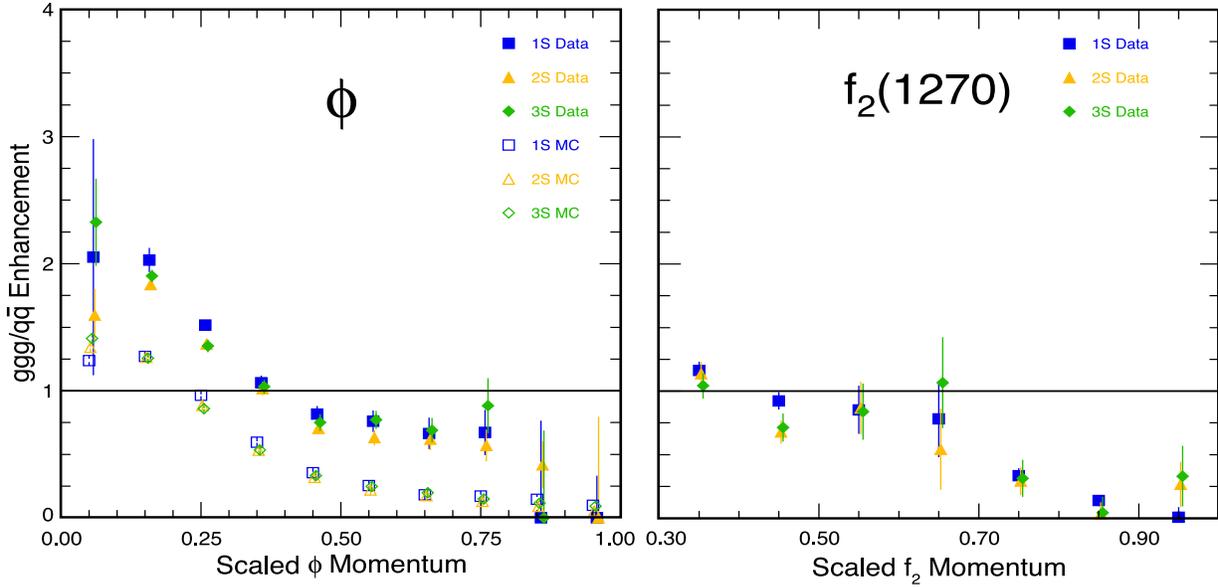}
     \caption{\label{fig:gggphi} (Left) Raw
enhancements for $ggg$ $\phi$ binned
according to scaled momentum.  Blue square (gold triangle, green diamond) symbols correspond to 
enhancements on the 1S (2S, 3S) resonance.  Closed symbols are data, 
open symbols are JETSET Monte Carlo. 
(Right) Enhancements for $f_2$(1270).}
\end{figure*}

Figure~\ref{fig:gggphi} shows $\phi$ enhancement results binned
according to scaled momentum.  Symbols are as above with blue square (gold triangle, green diamond)
corresponding to enhancements on the 1S (2S, 3S) 
resonance.  Closed symbols are data and open symbols are JETSET Monte Carlo.
Here, we have normalized
the resonant production at 9.46 GeV to the continuum production at
10.55 GeV. For $\phi$ production,
the lowest momentum bins for the resonance
are particularly sensitive to low-momentum kaon acceptance. 
Figure~\ref{fig:gggphi} also shows the f$_2$ enhancement results binned
according to scaled momentum.  The f$_2$ peak is not well-defined 
at low momentum (lowest two bins).  No Monte Carlo comparison
is presented since
our current Monte Carlo
event generator, by default, will not generate $f_2$
tensor particles.

\subsubsection{Particle Momentum-Integrated Enhancements}
\begin{figure*}
     \includegraphics[width=5in,height=5in]{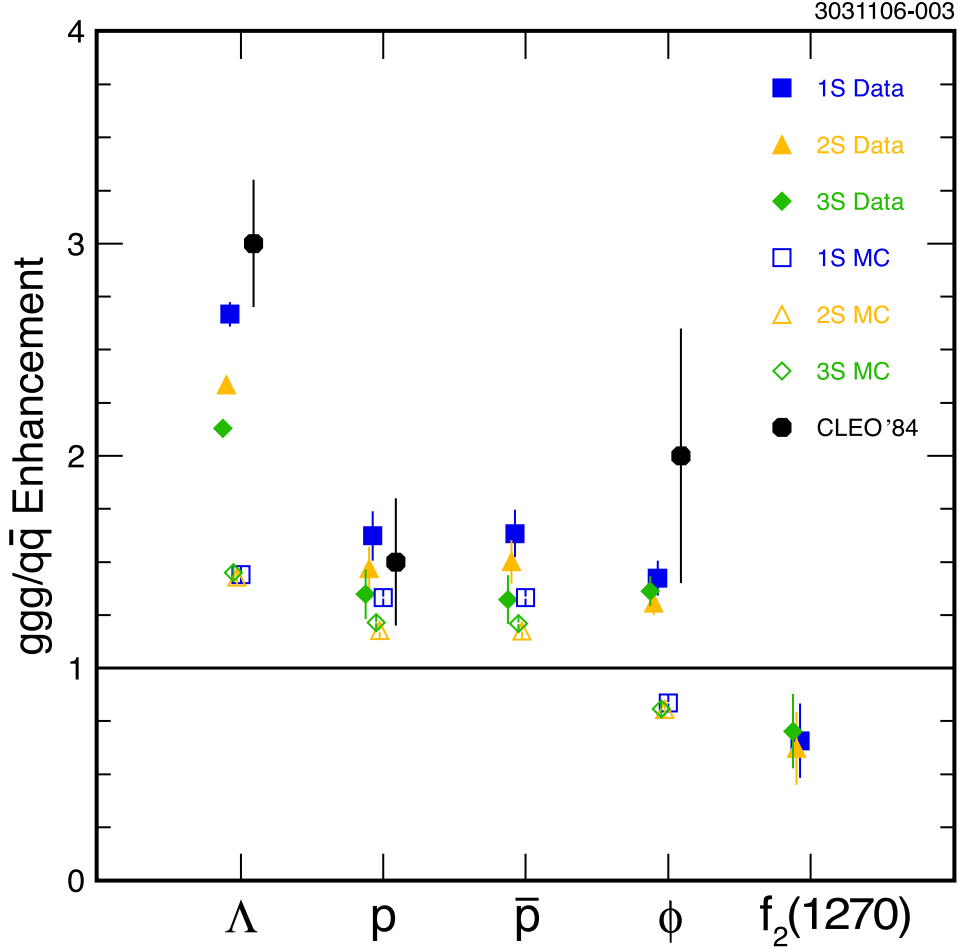}
     \caption{\label{fig:ggginteg} Compilation of momentum-integrated 
enhancements for $ggg$ events.  Blue square (gold triangle, green diamond) symbols correspond to 
enhancements on the 1S (2S, 3S) resonance.  Closed symbols are data, 
open symbols are JETSET Monte Carlo. 
Systematic errors and relative efficiencies have now been included
for this compilation. The CLEO84 study did not measure an
enhancement for $f_2(1270)$ and also only presented
a single enhancement for the sum of $p$ and $\overline{p}$.}
\end{figure*}

Figure~\ref{fig:ggginteg} shows the particle enhancements integrated over 
all momenta
for each particle, summarized numerically in Table~\ref{tab:summary} .  
We note that the baryons
($\Lambda$, $p$, $\overline{p}$) have enhancements greater than 1,
the $\phi$ meson enhancement is closer to unity, and the production
of the tensor $f_2$ is less than unity over our 
kinematic acceptance region. 
Our results are, in general, numerically consistent with the
prior CLEO-I analysis, albeit with considerably improved statistical
precision.

\subsection{gg$\gamma$ Enhancements with respect to $q{\overline q}\gamma$}

There are
sufficient CLEO III statistics
to present enhancements binned according to photon momentum, but
integrated over particle momenta
for $\Lambda$, $p$
and $\overline{p}$. For all particles, we also present
momentum-integrated enhancements.

\subsubsection{Baryon Enhancements}

\begin{figure*}
     \includegraphics[width=3.5in,height=3.5in]{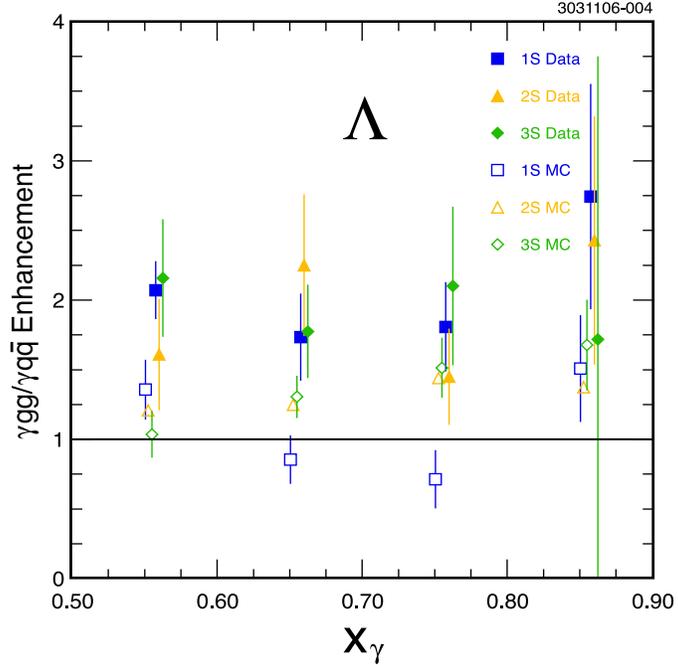}
     \caption{\label{fig:gggamlam} 
Raw enhancements for $gg\gamma\to\Lambda+X$ 
binned
according to scaled photon energy, integrated over
all $\Lambda$ momenta.
Blue square (gold triangle, green diamond) symbols correspond to 
enhancements on the 1S (2S, 3S) resonance.  Closed symbols are data, 
open symbols are JETSET Monte Carlo. }
\end{figure*}

\begin{figure*}
\includegraphics[width=6.4in,height=3.2in]{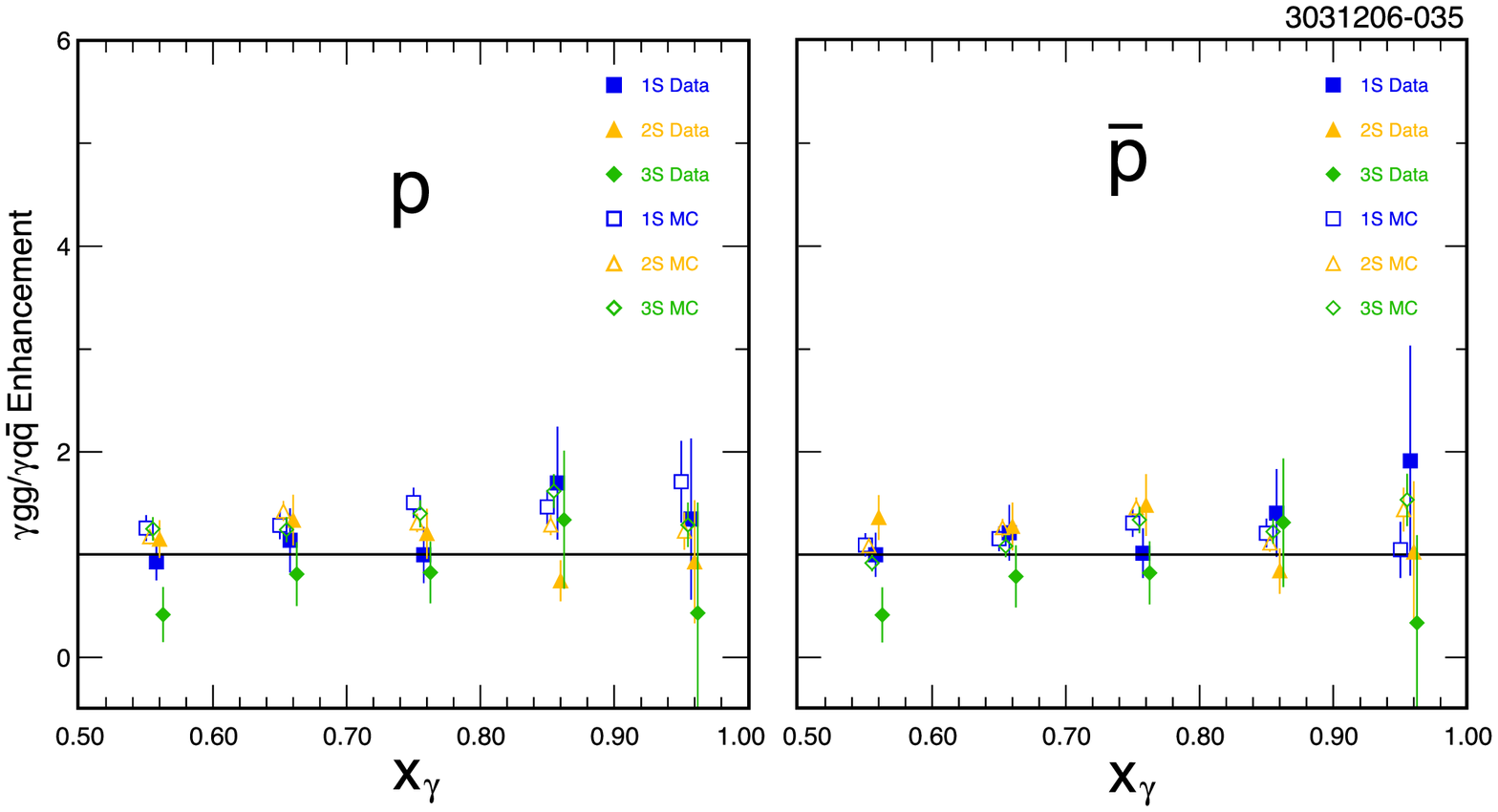}
     \caption{\label{fig:gggampro} (Left) 
Raw enhancements for $gg\gamma\to p+X$ 
binned
according to scaled photon energy, integrated over
all $p$ momenta.
Blue square (gold triangle, green diamond) symbols correspond to 
enhancements on the 1S (2S, 3S) resonance.  Closed symbols are data, 
open symbols are JETSET Monte Carlo.
(Right) Same for $\overline{p}$.}
\end{figure*}

Figure~\ref{fig:gggamlam} shows $\Lambda$ results binned according to scaled
photon momentum. 
For $\Lambda$'s,
as compared to the momentum-integrated $ggg$/$q \overline{q}$ 
enhancements, we observe a lower overall enhancement, on the order of 2
as opposed to 2.5--3 for the $\Upsilon$'s (Figure~\ref{fig:ggginteg}).  
Figure~\ref{fig:gggampro} shows $p$ and $\overline{p}$ 
enhancement results binned according
to scaled photon momentum. We note that $gg\gamma$/$q \overline{q}\gamma$
$p$ and $\overline{p}$ exhibit behavior similar to that of $\Lambda$'s.

\subsubsection{Photon Momentum-Integrated Enhancements}
\begin{figure*}
     \includegraphics[width=5in,height=5in]{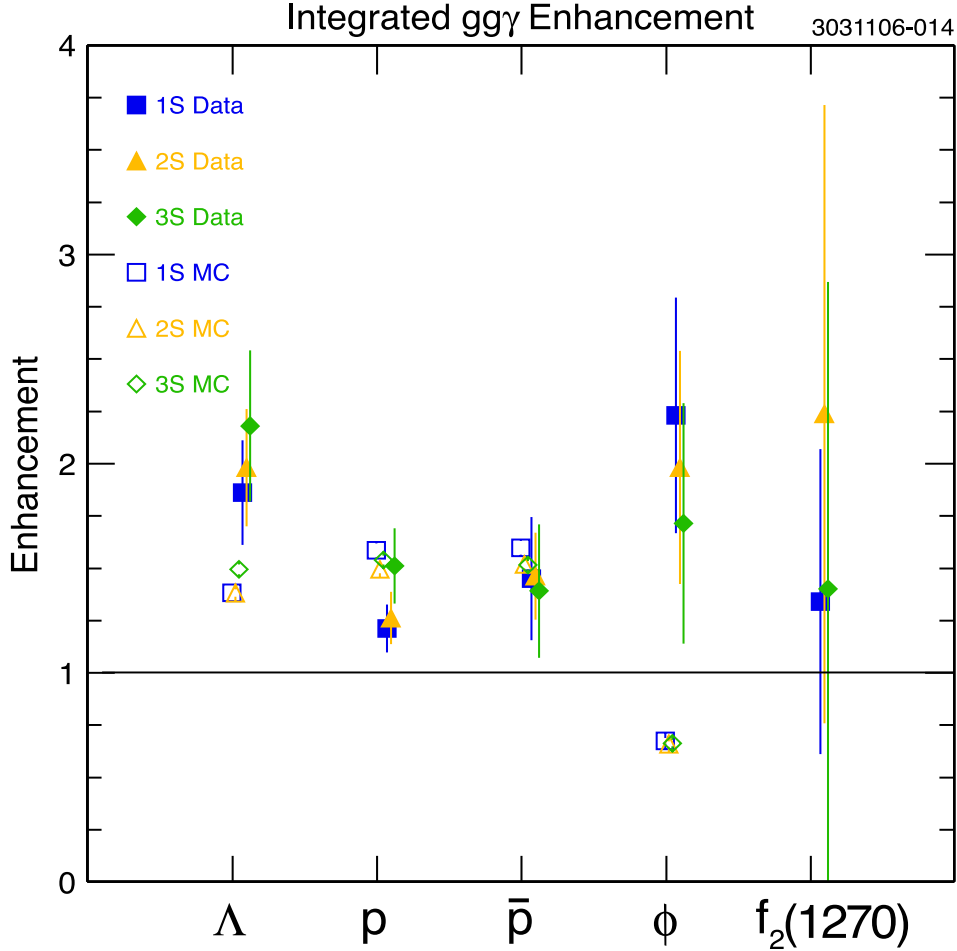}
     \caption{\label{fig:gggaminteg} 
Compilation of photon momentum-integrated 
enhancements for $gg\gamma$ events.  
Blue square (gold triangle, green diamond) symbols correspond to 
enhancements on the 1S (2S, 3S) resonance.  Closed symbols are data, 
open symbols are JETSET Monte Carlo. 
Systematic errors and relative efficiencies have now been included
for this compilation.}
\end{figure*}

Figure~\ref{fig:gggaminteg} shows the photon momentum-integrated enhancements
for each particle, summarized numerically in Table~\ref{tab:summary}.  
We note that all baryons
show enhancements lower than in the 3-gluon 
case (Figure \ref{fig:ggginteg}).

\section{Inclusive Proton Production in $\chi_{bJ}$ Decays}
Photon transitions of the $\Upsilon$(2S) and 
$\Upsilon$(3S) to the $\chi_b(')$ states allow
us to measure the baryon yields in $\chi_b(')$ decay, 
in association with a radiative
transition photon `tag'. Typical photon tag energies in this
case are of order 80-160 MeV. Due to the large
$\pi^0\to\gamma\gamma$ backgrounds to
such transition photons at these relatively low
photon energies, which compromises the
statistical power of such tags, 
the data permit only an extraction of the proton and antiproton
enhancements. Of particular interest is the proton yield
in $\chi_{b2}$ vs. $\chi_{b1}$ decays; the former is
expected to be dominated by decays via two gluons, the latter
is expected to be dominated by decays to $q{\overline q}$(g), with the gluon
expected to carry away very little momentum.

To ensure that photon-finding systematics largely cancel in the
ratio, and to also exclude possible contributions from initial
state radiation, we compare particle yields within the $\chi_{bJ}$
system directly rather than normalizing, e.g. relative to the
underlying $e^+e^-\to q{\overline q}(\gamma)$ continuum.
We first conduct a Monte Carlo study to determine the relative 
efficiency of reconstructing a J=2 transition
photon relative to J=1 event,
and also the efficiency when we require that a proton be found in
addition to the transition photon.
We compile statistics on the $\chi_b(')\to p({\overline p})+X$ 
analyses, separately for J=0/J=1 and for J=2/J=1. For the latter,
the overlap of the two observed photon signals results in a
highly correlated event yield for the two transitions.
We correspondingly extract this
ratio from a signal fit to a double Gaussian plus a smooth background.
For the former, we simply fit two separate signal Gaussians directly.
We find that the efficiency for reconstructing photon-proton
correlations in $\chi_{b2}\to gg$
decays is approximately 95\% that for photon-proton correlations
in $\chi_{b1}\to q{\overline q}$(g) events.

To check the sensitivity to our particle identification
criteria, we have compared results using
very tight proton identification 
requirements (with a reduction in efficiency by more than 50\%)
vs. the 'standard' loose proton identification
criteria used above. We obtain a comparable
correction factor for the J=2/J=1 event yields
using more restrictive particle identification criteria.

Results are presented in Table \ref{tab:chib}. 
We note that the
observed enhancements are, again, smaller than those
observed in comparing three-gluon fragmentation from the $\Upsilon$ resonance
with $q{\overline q}$ fragmentation.

\begin{table}
\caption{Summary of inclusive proton (and antiproton) results for
$\chi_{bJ}$ decays. For checks of internal consistency,
data have been separated into sub-samples, labeled with capital
Roman letters. For J=2 relative to
J=1, e.g., the scale of systematic uncertainties is set by
the constancy of the value across sub-samples
collected in different running periods (r.m.s.$\sim$0.03),
the magnitude of relative efficiency corrections ($\sim$0.05)
and the consistency of results obtained using different particle
identification criteria. For summed results (labeled ``all''), the
second error shown is the systematic error.}
\label{tab:chib}
\begin{tabular}{|c|c|c|c|} \hline
Dataset & particle & ($\chi_{b2}\to p+X$)/ & 
($\chi_{b0}\to p+X$) \\
        & identification & ($\chi_{b1}\to p+X$)/ & ($\chi_{b1}\to p+X$) 
\\ \hline
 
(3S A) & loose & $1.116\pm0.017$ & $1.19\pm0.046$ \\ 
 
(3S B) & loose & $1.080\pm0.016$ & $1.00\pm0.034$ \\ 
 
(3S C) & loose & 

$1.086\pm0.011$ &

$1.054\pm0.047$ \\ 
 
(3S D) & tight & 

$1.103\pm0.027$ &

$1.091\pm0.097$ \\ \hline
3S, all & & $1.109\pm0.007\pm0.040$ & $1.082\pm0.025\pm0.060$ \\ \hline

(2S A) & tight & 1.066$\pm$0.028 & 1.03$\pm$0.13 \\

(2S B) & loose & 1.075$\pm$0.018 & 

$1.36\pm0.15$ \\

(2S C) & loose & 1.076$\pm$0.017 & 

0.99$\pm$0.11 \\

(2S D) & loose & 1.065$\pm$0.015 & 

1.06$\pm$0.11 \\ \hline

(2S B) & tight & 1.076$\pm$0.047 & 

$1.39\pm0.28$ \\

(2S C) & tight & 1.039$\pm$0.040 & 

1.17$\pm$0.22 \\

(2S D) & tight & 1.024$\pm$0.035 & 

0.88$\pm$0.20 \\ \hline
2S, all & & $1.068\pm0.010\pm0.040$ & $1.11\pm0.15\pm0.20$ \\ \hline

Monte Carlo (3S A) & loose & $1.057\pm0.016$ & $1.030\pm0.072$ \\

Monte Carlo (3S A) & tight & $1.034\pm0.015$ & $1.042\pm0.066$ \\

Monte Carlo (3S B) & tight & $1.041\pm0.013$ & $1.051\pm0.049$ \\ \hline
MC, 3S all sets & & $1.043\pm0.008$ & $1.043\pm0.036$ \\ \hline

Monte Carlo (2S A) & tight & 1.052$\pm$0.014 & 1.121$\pm$0.058 \\

Monte Carlo (2S A) & loose & 1.043$\pm$0.015 & 1.076$\pm$0.061 \\ \hline
MC, 2S all sets & & $1.046\pm0.010$ & $1.061\pm0.025$ \\ \hline
\end{tabular}
\end{table}

\message{In Monte Carlo, there is, 
therefore no ``gluonic baryon enhancement'' as measured relative to
non-protons --
the ratio of 
$((\chi_{b2}\to p+X/\chi_{b1}\to p+X)/
(\chi_{b2}\to {\not p}+X/\chi_{b1}\to {\not p}+X))$=1.004
is statistically consistent with unity, as is the corresponding
ratio for 
$((\chi_{b0}\to p+X/\chi_{b1}\to p+X)/
(\chi_{b0}\to {\not p}+X/\chi_{b1}\to {\not p}+X))$ (=0.987).
In data, the corresponding ratio is:
$((\chi_{b2}\to p+X/\chi_{b1}\to p+X)/
(\chi_{b2}\to {\not p}+X/\chi_{b1}\to {\not p}+X))$=0.993/0.981.}

\section{Cross-Checks and Systematics}
\begin{figure*}
\includegraphics[width=6.4in,height=3.2in]{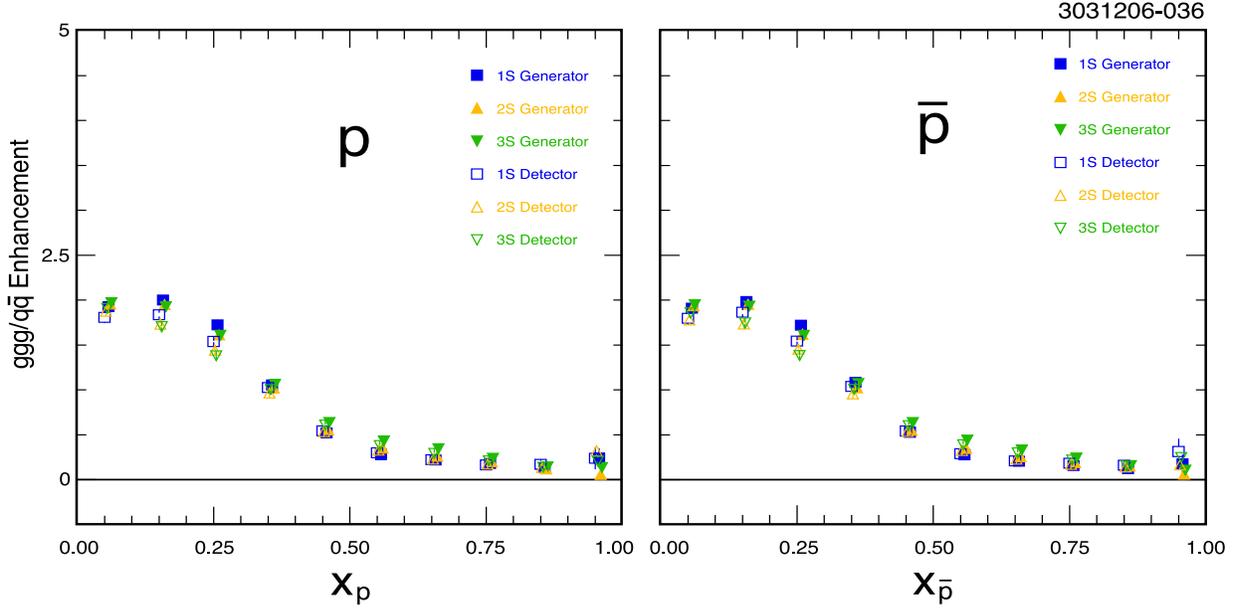}
     \caption{\label{fig:gggprodg} (Left) Scaled momentum binned 
enhancements for $ggg\to p+X$ at generator level and 
after detector simulation.  Blue square (gold triangle, green 
inverted triangle) symbols 
correspond to enhancements on the 1S (2S, 3S) resonance.  Closed (open)
symbols are generator (detector) level Monte Carlo enhancements. (Right)
Same for $\overline{p}$. }
\end{figure*}

\begin{figure*}
\includegraphics[width=6.4in,height=3.2in]{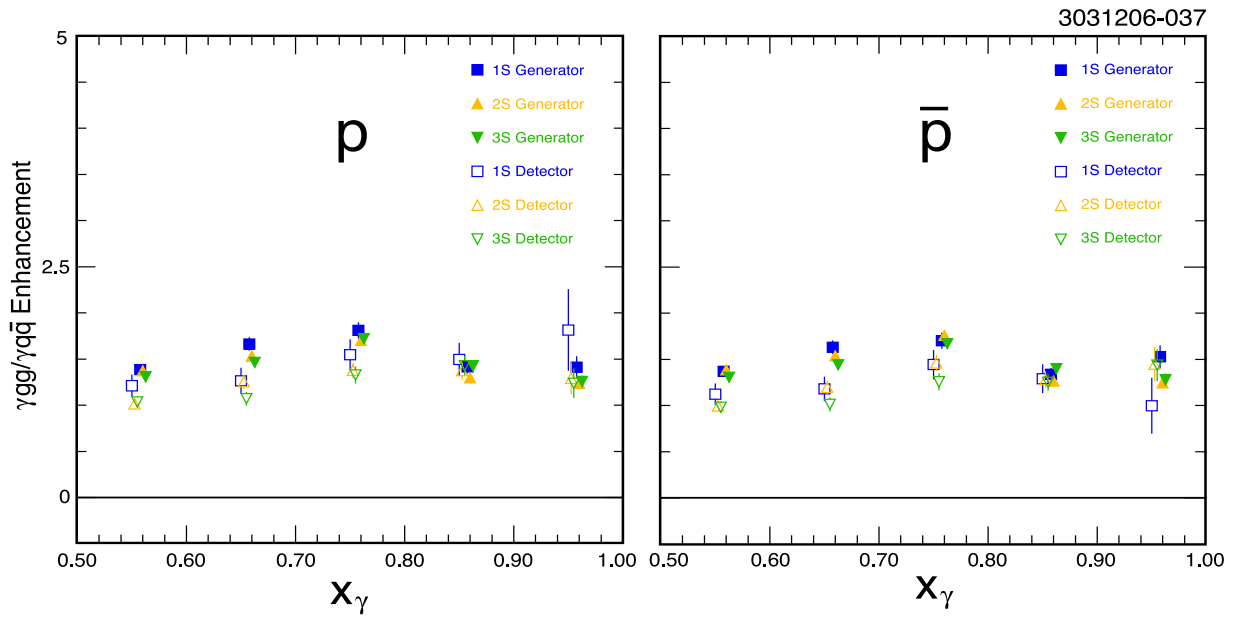}
     \caption{\label{fig:gggamprodg} (Left) Scaled momentum binned 
enhancements for $gg\gamma$ decays to $p$ at generator level and
after detector simulation.  Blue square (gold triangle, green inverted
triangle) symbols 
correspond to enhancements on the 1S (2S, 3S) resonance.  Closed (open)
symbols are generator (detector) level Monte Carlo enhancements. (Right)
Same for $\overline{p}$. }
\end{figure*}

In order to verify our procedures and probe possible
systematic uncertainties, two primary cross-checks were employed.
We first compare the Monte Carlo enhancements at the
event generator-level with those determined after the generated
events are processed through the full
CLEO-III detector simulation (``detector-level''), as a function of momentum.
In general, these enhancements will differ for several reasons, including
differences in:
a) the efficiencies for finding recoil particles in $q{\overline q}\gamma$ 
vs. $gg\gamma$ events resulting from angular distribution,
event multiplicity, and particle momentum differences,
b) event selection efficiencies, c) $\pi^0$
contamination levels, and d) recoil center-of-mass
discrepancies between the continuum data under
the $\Upsilon$(1S) resonance vs. the below-$\Upsilon$(4S)
continuum.
In cases where the generator-level and detector-level
enhancements are statistically inconsistent with
each other at the $2\sigma$ level,
we use the ratio between the generator-level
and detector-level enhancements as a correction factor and
take half of the amount by which this
correction deviates from unity as an estimated systematic error.
(Note that these corrections have already been incorporated
into the results presented in Figures and \ref{fig:ggginteg}
and \ref{fig:gggaminteg}). Figures \ref{fig:gggprodg} 
and \ref{fig:gggamprodg} shows the comparison of $p$
enhancements determined at the event-generator
vs. post-detector-simulation levels of Monte Carlo simulation.
Integrated over momentum, typical
corrections are typically of order 10\%.

In addition to the comparison of generator vs. 
detector-level enhancements,
we have made an
additional (largely redundant) check of possible biases due to 
non-direct photons resulting from, e.g.,
$\pi^0\to\gamma\gamma$, $\eta\to\gamma\gamma$, etc. Using
Monte Carlo simulations, we compare the enhancements obtained
using direct-photons only, compared with the enhancements
obtained when we include all Monte Carlo photons which pass
our photon selection, independent of parentage. 
Integrated over momentum, this again constitutes a $\sim$5\% effect,
and is conservatively included as an additional (in quadrature) systematic
error.

To test the sensitivity of our analysis procedures across
different running periods, we have calculated the enhancements
for photon-tagged $\Upsilon$(4S) on-resonance events vs.
photon-tagged below-$\Upsilon$(4S) continuum events, spanning
the full CLEO-III data set. Since $\Upsilon$(4S)$\to$B\=B $\sim$ 100\%,
we expect that any event having a photon with $z_\gamma>0.5$ is
a continuum event. Hence, the calculated enhancement should be
zero. In all cases, save for $\overline{p}$, we find
good agreeement between the below-4S continuum particle 
yields per photon tag, and the on-4S particle
yields per photon tag. For $\overline{p}$,
we find deviations from the null expectation at the
level of $\approx$5--7\%, and incorporate these deviations
(bin-by-bin in momentum) into our total systematic error
for that particular case. For the case of the broad
$f_2$ resonance, sensitivity to our parameterization of the
smooth background also contributes a non-negligible systematic
uncertainty.

We note that most systematic errors cancel in our ratios.
To summarize the sources of systematic uncertainties,
the largest components are the efficiency bias ($\sim$10\%),
the non-direct photon background ($\sim$5\%),
and the run-dependence of our result, as measured by our
`null' check ($\sim$6\%),
all added in quadrature.
Our results with statistical and systematic errors are listed in 
Table~\ref{tab:summary}.
The statistical uncertainties in the data are typically of order
10\%, with the 
exception of
$gg\gamma/q{\overline q}\gamma$ for $\phi$ and $f_2$, which are
of poorer statistical quality.

\section{Discussion and Summary}
We have, for the first time, measured the momentum-dependent
ratio of baryon and meson production in gluon vs. quark
fragmentation at $\sqrt{s}\sim$10 GeV.
After reproducing the previously measured per-event baryon
production rates in three-gluon decays of the
$\Upsilon$(1S) resonance relative to the underlying
continuum, we have extended that study to include the
other narrow $\Upsilon$ resonances and, with higher
statistics, now explicitly examine
the momentum dependence of the enhancements for
all these states.
Integrated over momentum, we 
observe approximately 5\% (10\%) lower baryon production
per-event for $\Upsilon$(2S) ($\Upsilon$(3S)) decays compared
to the vector ground state.
Nevertheless, the per-event production of $\Lambda$'s
for each of the narrow $\Upsilon$ resonances is observed
to be greater than twice that of continuum fragmentation at
the same center-of-mass energy.

We additionally compare, for the first time, particle production in
two-gluon vs. quark-antiquark fragmentation. We find,
in particular, that baryon production (per event) in two-gluon decays is 
somewhat 
smaller 
($\sim$20\% for baryons)
than that observed in three-gluon decays. For
$\Lambda$ production, we still observe a significant 
($\sim\times$2) enhancement
in two-gluon fragmentation relative to quark-antiquark fragmentation,
although the excess enhancement for $p$ is $\le$10\%.
For $p$, which represent our
highest-statistics sample, our results are inconsistent
with a model where baryon production in gluon fragmentation
is only a function of the available center-of-mass energy; clearly,
the number of fragmenting partons is also important, although our
measured enhancements fall short of the expectations from a naive
independent fragmentation model. 
Our results, for all measured integrated enhancements are presented
in Table \ref{tab:summary}. 
\begin{table}[h]
\begin{tabular}{c|c|c|c|c} \hline
Particle & $gg\gamma/q{\overline q}\gamma$ data & $gg\gamma/q{\overline q}\gamma$ MC & $ggg/q \overline{q}$ data & $ggg/q \overline{q}$ MC \\ \hline
$\Lambda$ (1S) & $1.86\pm0.25\pm0.03$ & $1.38\pm0.039$ & $2.668\pm0.027\pm0.051$ & $1.440\pm0.003$ \\
$\Lambda$ (2S) & $1.98\pm0.27\pm0.08$ & $1.38\pm0.018$ & $2.333\pm0.019\pm0.021$ & $1.428\pm0.002$ \\
$\Lambda$ (3S) & $2.18\pm0.36\pm0.02$ & $1.49\pm0.023$ & $2.128\pm0.021\pm0.010$ & $1.450\pm0.002$ \\ \hline
p (1S) & $1.21\pm0.11\pm0.03$ & $1.582\pm0.034$ & $1.623\pm0.014\pm0.116$ & $1.331\pm0.005$ \\
p (2S) & $1.26\pm0.11\pm0.06$ & $1.495\pm0.018$ & $1.469\pm0.011\pm0.103$ & $1.177\pm0.003$ \\
p (3S) & $1.51\pm0.17\pm0.06$ & $1.53\pm0.021$ & $1.348\pm0.013\pm0.116$ & $1.214\pm0.003$ \\ \hline
${\overline p}$ (1S) & $1.45\pm0.14\pm0.26$ & $1.589\pm0.034$ & $1.634\pm0.014\pm0.111$ & $1.333\pm0.005$ \\
${\overline p}$ (2S) & $1.46\pm0.12\pm0.17$ & $1.513\pm0.018$ & $1.500\pm0.011\pm0.102$ & $1.175\pm0.003$ \\
${\overline p}$ (3S) & $1.39\pm0.17\pm0.27$ & $1.51\pm0.020$ & $1.323\pm0.013\pm0.115$ & $1.210\pm0.003$ \\ \hline
$\phi$ (1S) & $1.78\pm0.49\pm0.08$ & $0.673\pm0.013$ & $1.423\pm0.051\pm0.065$ & $0.836\pm0.003$ \\
$\phi$ (2S) & $1.73\pm0.52\pm0.06$ & $0.658\pm0.012$ & $1.308\pm0.041\pm0.041$ & $0.805\pm0.001$ \\
$\phi$ (3S) & $1.87\pm0.81\pm0.06$ & $0.662\pm0.015$ & $1.355\pm0.054\pm0.047$ & $0.808\pm0.002$ \\ \hline
$f_2$(1270) (1S) & $1.34\pm0.84\pm0.15$ ($<2.74$) & $-$ & $0.658\pm0.058\pm0.175$ & $-$ \\ 
$f_2$(1270) (2S) & $2.22\pm1.53\pm0.20$ ($<4.68$) & $-$ & $0.621\pm0.094\pm0.171$ & $-$ \\
$f_2$(1270) (3S) & $1.41\pm1.48\pm0.10$ ($<3.87$) & $-$ & $0.702\pm0.104\pm0.175$ & $-$ \\ \hline
\end{tabular}
\caption{Numerical summary of momentum-integrated
enhancement results.  Second and third columns show results from Figure~\ref{fig:gggaminteg},
the photon momentum integrated $gg\gamma$/$q\overline{q}\gamma$ study.
Fourth and fifth columns show results from Figure~\ref{fig:ggginteg}, the particle momentum
integrated $ggg$/$q\overline{q}$ study.  MC refers to JETSET Monte Carlo.  Data errors are statistical
and systematic; MC errors are purely statistical. For the
$f_2$, we present 90\% C.L. upper limits, given the poor statistical
significance of the $gg\gamma/q{\overline q}\gamma$ enhancements.}
\label{tab:summary}
\end{table}

Although event generators such as JETSET have had tremendous success in
describing the gross details of particle production in $e^+e^-$ collisions,
our study indicates that there may still be considerable tuning needed
at the single-particle yield level.

\section{Acknowledgments}
We gratefully acknowledge the effort of the CESR staff
in providing us with excellent luminosity and running conditions.
D.~Cronin-Hennessy and A.~Ryd thank the A.P.~Sloan Foundation.
This work was supported by the National Science Foundation,
the U.S. Department of Energy, and
the Natural Sciences and Engineering Research Council of Canada.

\end{document}